\documentclass[letterpaper,cmfonts,draft,
  ]
  {aipproc}

\layoutstyle{6x9}

\begin{document}

\def\beq{\begin{equation}}
\def\eeq{\end{equation}}
\def\beqa{\begin{eqnarray}}
\def\eeqa{\end{eqnarray}}
\def\d{{\rm d}}
\def\ttimes{{\scriptstyle \times}}
\def\half{{\textstyle {1\over2}}}

\title{Small quark stars in the chromodielectric model}

\author{M. Malheiro}{
  address={Instituto de F\'\i sica, Universidade Federal Fluminense,
Av. Litor\^anea, 24210-310 Niter\'oi, Brazil}
}
\author{E.O. Azevedo}{
  address={Instituto de F\'\i sica, Universidade Federal Fluminense,
Av. Litor\^anea, 24210-310 Niter\'oi, Brazil}
}
\author{L.G. Nuss}{
  address={Instituto de F\'\i sica, Universidade Federal Fluminense,
Av. Litor\^anea, 24210-310 Niter\'oi, Brazil}
}
\author{M. Fiolhais}{
  address={Departamento de F\'\i sica and Centro de F\'\i sica Computacional, 
Universidade de Coimbra, P-3004-516 Coimbra, Portugal}
}

\author{A. Taurines}{
  address={Instituto de F\'\i sica, Universidade Federal do 
Rio Grande do Sul CP 15051, 91501-970 Porto Alegre, Brazil}
}

\begin{abstract}
Equations of state for strange quark matter in beta equilibrium 
at high densities are used to investigate the structure (mass and radius) 
of compact objects. The chromodielectric model is used as a general framework
for the quark interactions, which are mediated by chiral mesons, 
$\sigma$ and $\vec \pi$, 
and by a confining chiral singlet dynamical field, $\chi$.
Using a quartic potential for $\chi$, two  
equations of state for the same set of model parameters are obtained, 
one with a minimum at around the nuclear matter density
$\rho_0$ and the other at $\rho \sim 5 \, \rho_0$. Using the latter equation of state 
in the Tolman-Oppenheimer-Volkoff 
equations we found solutions corresponding to compact objects
with $R\sim 5 - 8$~km and $M\sim M_\odot$.
The phenomenology of recently discovered X-ray sources is compatible with the 
type of quark stars that we have obtained.

\end{abstract}

\maketitle


\section{Introduction}
Various effective models using quarks as fundamental dynamical fields, 
originally designed for the nucleon, have also been 
successfully used to describe infinite quark matter, 
and the resulting equations of state
(EOS's) have been applied to investigate the structure of compact
stars [1--6].

The chromodielectric model (CDM)~[7--9], 
for example,  provides a reasonable phenomenology 
for the nucleon~\cite{neuber,drago1} and also allows us to 
obtain EOS's for dense quark matter. The model
yields soliton solutions representing single baryons with 
three quarks dynamically confined by a scalar field, 
$\chi$, whose quanta can be assigned to $0^{++}$ glueballs.  
When it is applied to quark matter in two or three flavors~[12--14] 
the resulting EOS turns out to be relatively soft at large densities.

 Using a quadratic potential for the $\chi$ field, Drago et al.~\cite{drago,drago3} 
applied the CDM to describe the inner part of neutron 
stars,
obtaining masses in the range $1-2 M_\odot$ and radii of the order 10 km or 
higher, with a small hadron crust of  2 km.  
In this work we consider an extension of the model used in 
Ref.~\cite{drago}, taking 
quartic instead of quadratic potentials. In addition to the 
structures found by Drago et al.,
the quartic model predicts another type of  
compact objects made out of quarks only, smaller and denser than neutron stars.

From the observational point of view, the recent discovery of 
X-ray sources, by the Hubble and Chandra telescopes, 
increased the plausibility that these sources might be
strange quark stars~[17--19].   
In particular, 
the isolated compact object RX J1856.5-3574 with a 
small radius does not show evidence 
of spectral lines or edge features~\cite{pons,drake}, 
reinforcing the conjecture for the existence of 
stars made out of strange matter. 
The phenomenology of these objects seems to 
be compatible with the small and dense quark stars reported in this work.

\section{The model}
\subsection{CDM Lagrangian}
\begin{figure}[bht]
  \includegraphics[width=13cm]{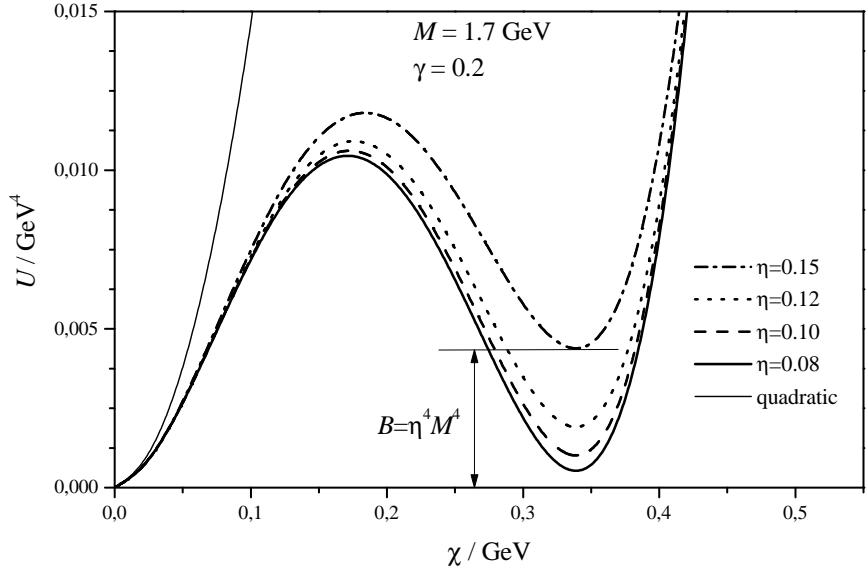}
\caption{Quartic potential, Eq. (\ref{uchi}), for fixed $M$ and $\gamma$, 
in dependence of $\chi$.
 For comparison, the quadratic potential
 $(\gamma \rightarrow \infty)$ is also shown. The value of the 
potential at the second minimum can be 
interpreted as a bag pressure, $B$. }
\label{figstar1}
\end{figure}

We write the CDM Lagrangian in the form~[8--10] 
\begin{equation}
  \mathcal{L} = \mathcal{L}_q + \mathcal{L}_{\sigma,\pi} 
    +  \mathcal{L}_{q-\mathrm{meson}} +  \mathcal{L}_\chi\;,
  \label{langrangian}
\end{equation}
where
\begin{equation}
  \mathcal{L}_q = \mathrm{i}\bar{\psi}\gamma^\mu \partial_\mu\psi \;,
\qquad
  \mathcal{L}_{\sigma,\pi} =
  \half\partial_\mu\hat{\sigma}\partial^\mu\hat{\sigma}
  + \half\partial_\mu\hat{\vec{\pi}}\cdot\partial^\mu\hat{\vec{\pi}} 
  - {W}(\hat{\vec{\pi}},\hat{\sigma})\;,
  \label{langrangian1}
\end{equation}
and ${W}(\hat{\vec{\pi}},\hat{\sigma})$ is the
Mexican hat potential. In the $u,\, d$ sector the quark-meson interaction
is described by
\begin{equation}
    \mathcal{L}_{q-{\mathrm{meson}}} = {g\over\chi}\, \bar{\psi}
    (\hat{\sigma}+\mathrm{i}\vec{\tau}\cdot\hat{\vec{\pi}}\gamma_5)
    \psi\;.
   \label{langrangian2}
\end{equation}
The last term in (\ref{langrangian})
contains the kinetic and the potential piece for the $\chi$-field:
\begin{equation}
  \mathcal{L}_\chi =  
  \half\partial_\mu\hat{\chi}\,\partial^\mu\hat{\chi}
  -  U(\hat{\chi})\, .
  \label{langrangian3}
\end{equation}
The potential term is 
\begin{equation}
U(\chi)= {1 \over 2} M^2 \chi^2 \left[  1+\left( 
{8 \eta^4 \over \gamma^2}-2 \right) {\chi \over \gamma M} +  
\left( 1- {6 \eta^4 \over \gamma^2} \right) {\chi^2 \over 
(\gamma M)^2} \right] \, ,
\label{uchi}
\end{equation}
where $M$ is the $\chi$ mass. The parameterization used in 
(\ref{uchi}) allows for a 
physically meaningful interpretation of the parameters 
$\gamma$ and $\eta$: $U(\chi)$ has a 
global minimum at $\chi=0$ and a local one at $\chi=\gamma M$, 
and  $U(\gamma M)=\eta^4 M^4$ (see Fig. \ref{figstar1}).
 The height of the local minimum, $B=(\eta M)^4$, is interpreted 
 as a ``bag pressure" and this is used to fix the parameters in $U(\chi)$. 
 Assuming  the wide range $0.150 \le B^{1/4} \leq 0.250$ GeV,
 one has $0.08\le \eta \le 0.15$, using $M=1.7$~GeV. 
We note that  $\gamma$  is not 
 a free parameter since the quartic term of $U(\chi)$ 
must be positive and the cubic term negative, which implies 
 $\gamma^2\ge 6 \eta^4$. In the soliton sector of the model, 
best nucleon properties are obtained for $G=\sqrt{g M}\sim
0.2$~GeV (only $G$ matters for the nucleon properties) and 
we keep that $G$ in our quark matter calculations.       

In order to study strange quark matter, we add to the interaction Lagrangian (\ref{langrangian2}) the 
term~\cite{birsegovern} 
\begin{equation}
    \mathcal{L}_{s-{\mathrm{meson}}} = {g_s \over\chi}\, \bar{\psi}_s
    \psi_s\;.
   \label{langrangian4}
\end{equation}
accounting for 
the coupling between the strange quark and the  $\chi$ field. 

\subsection{Strange quark matter}

In order to study strange quark matter in beta equilibrium, 
an electron gas must also be considered. The mean-field
 energy per unit volume for strange 
quark matter in the CDM  (plus electrons) is given by
\begin{eqnarray}
\varepsilon &=& \alpha
\sum_{f=u,d}\int_{0}^{k_{f}}\frac{d^3k}{(2\pi)^3}\sqrt{k^2+m_f(\sigma,\chi)^2}
+ \alpha
\int_{0}^{k_{s}}\frac{d^3k}{(2\pi)^3}\sqrt{k^2+m_s(\chi)^2}
\nonumber \\ &+& 2
\int_{0}^{k_{e}}\frac{d^3k}{(2\pi)^3}\sqrt{k^2+m_e^2}
+{ U}(\chi)+{ m _\sigma^2 \over 8 \, f_\pi^2}(\sigma^2-f_\pi^2)^2,
\label{eov2}
\end{eqnarray}
where the first two terms refer to quarks and the third one 
to the electrons, all described by plane waves.
 The degeneracy factor is  $\alpha=6$ (for spin and color). 
The last term is the Mexican hat potential
 (with $\vec\pi=0$ and $f_\pi=93$~MeV). The $k_i$ in (\ref{eov2}) 
are the Fermi momenta of quarks and electrons. 

The quark masses in (\ref{eov2}) are \cite{birsegovern}: 
$m_{u,d} = {g_{u,d}\, \sigma}/{(\chi \, f_\pi)} $ and $m_{s} =
{g_{s}}/{\chi}$ 
with the coupling constants given by
$g_u= g \, (f_\pi+\xi_3)$, $g_d= g \, (f_\pi-\xi_3)$ and $g_s=g\, (2 f_k-f_\pi)$   
[$\xi_3=-0.75$ MeV, $f_K=113$ MeV]. 

\begin{figure}[bht]
  \includegraphics[width=9.cm]{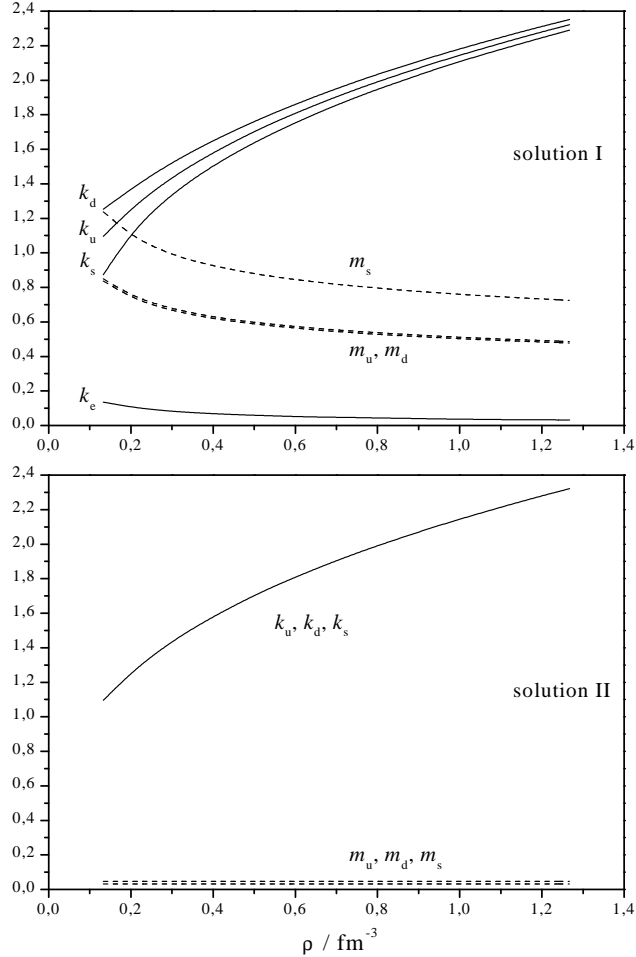}
\caption{Fermi momenta (solid lines) and quark masses 
(dashed lines) for solution I and II in dependence of 
the baryon density. For solution II the quark masses and the 
electron Fermi momentum almost vanish. 
The model parameters are $g=0.023$~GeV, $M=1.7$~GeV, $\gamma=0.2$ 
and $\eta=0.12$. The vertical scales are in fm$^{-1}$.}
\label{figstar2}
\end{figure}

A variational principle applied to the energy density, Eq.~(\ref{eov2}), 
leads to two gap equations for $\sigma$ and $\chi$. 
In the interior of a compact star the matter should 
satisfy both the electrical charge 
neutrality and chemical equilibrium.
These conditions should supplement the gap equations, 
and altogether we have a system
of six algebraic equations to solve at each baryon density
$\rho= \left( \rho_u+\rho_d+\rho_s \right)/3$
[here, $\rho_i=\alpha k_i^3 / (6\pi)^2$ stand for each flavor density]. 
The solution of the system of equations are 
the meson fields, $\sigma$ and $\chi$, and the Fermi momenta, 
$k_u$, $k_d$,  $k_s$ and $k_e$.
For the same set of model
parameters we found two stable solutions, 
hereafter denoted by I and II, shown in Fig.~\ref{figstar2}.

For both solutions $\sigma$  is always close to $f_\pi$. 
In solution I, the $\chi$ field is a slowly increasing function of
 the density, remaining always smaller than $\sim 0.05$~GeV. For such a small
$\chi$, the quartic potential and the quadratic potential are indistinguishable (see Fig. \ref{figstar1}), thus, 
in practice, solution I corresponds to the one obtained and used by Drago et al.
\cite{drago} in the framework of the quadratic potential. 
Due to the smallness of the $\chi$ field, quark
masses are 
large  and the system is in a chiral broken phase. 
The solution II exhibits a large confining field, 
$\chi\sim \gamma M$ (local minimum of $U$), independent of the density.
The resulting quark masses are similar for the three flavors 
and very close to zero (chiral restored phase). Therefore,
the chemical potentials in solution II are dominated by the 
Fermi momentum contribution, 
  $\mu_u\simeq\mu_d\simeq\mu_s$ and $\mu_e\simeq0$, 
i.e. in solution II there are almost no electrons.   
Besides solutions I and II, there is an additional 
{\em unstable} solution corresponding to $\chi\sim \gamma M /2$ 
[local maximum of $U(\chi$)].

\subsection{Equations of state}

 The energy per baryon number as a function of  the baryon 
density (EOS) is readily evaluated for each solution. 
We stress that EOS-I is not sensitive to $\gamma$ and $\eta$ 
(since $\chi$ is small), just depends on $G$. The saturation 
density occurs at a relatively low density
 ($\rho=1.2-1.6 \, \rho_0$) and the behavior of EOS-I at 
intermediate densities, is similar to the hadronic EOS's (see
Ref.~\cite{drago2} for the two flavors sector).  
 The EOS-II is also insensitive to $\gamma$, but does depend on 
$\eta$ [in fact, the dependence is on $(\eta M)^4$, 
as we have already pointed out]: the energy per baryon number 
increases with $\eta$ and so does the saturation density. 
For $\eta\sim 0.12$ the saturation density is $\rho\sim 5 \rho_0$ 
and the energy per baryon number is some 20\% 
higher than for solution I at its saturation density.

\begin{figure}[hbt]
  \includegraphics[width=16cm]{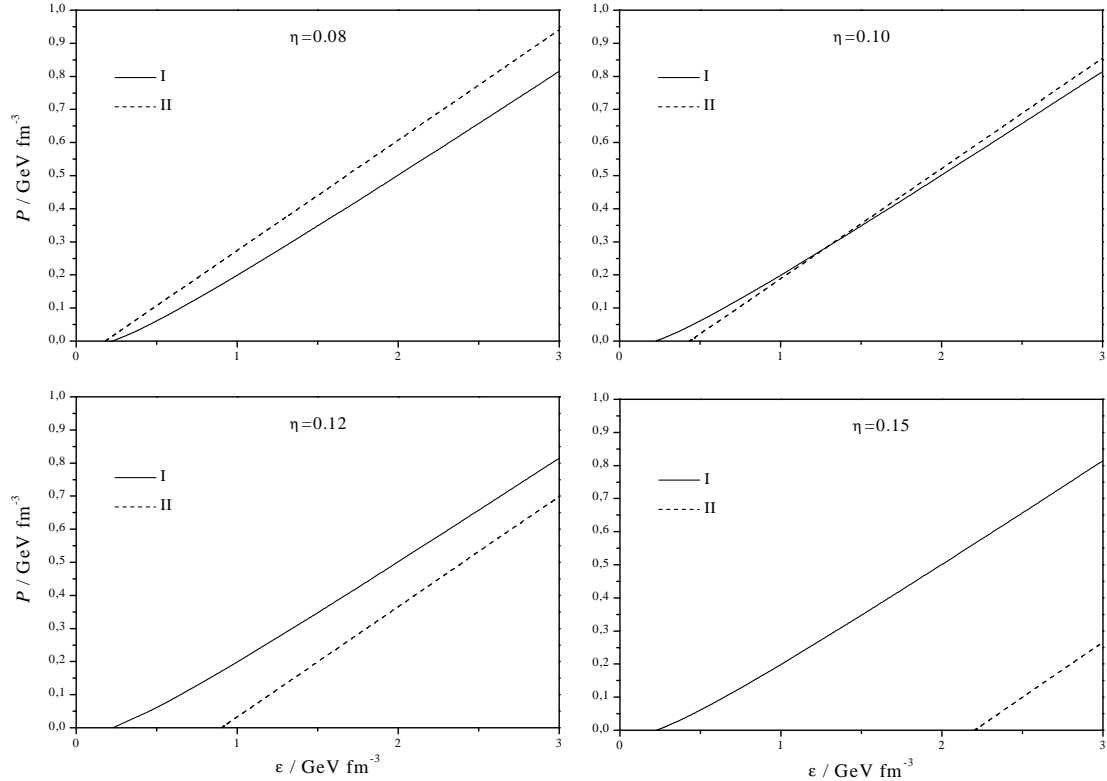}
\caption{Pressure versus energy density for the two types of 
solution for various parameters $\eta$. ]
The other model parameters are $M=1.7$~GeV, $g=0.023$~GeV and $\gamma=0.2$.}
\label{figstar3}
\end{figure}

In  Fig. \ref{figstar3} we present the EOS's in the form of pressure 
$P=\rho^2 {\partial \ \over \partial \rho}\left( {\epsilon \over \rho} \right)$ 
versus energy density plots for solutions I and II and 
various $\eta$.  It is worth noticing that our results with two 
distinct EOS predicted by the CDM,
 are consistent with the results from perturbative QCD: it is 
remarkable indeed that, for $\eta\sim0.12$ the 
CDM reproduces accurately the EOS's recently obtained in a 
perturbative QCD calculation 
(compare our third panel of Fig.~\ref{figstar3} with figure 
1 of Ref.~\cite{fraga}).

Regarding energetics, both phases are almost degenerated at 
high densities and have similar shapes in the $P \ttimes \epsilon$
plane even at intermediate densities (or energy densities) 
in the narrow range  $0.1 \le \eta \le 0.12$.  
In that region of $\eta$, one solution is not clearly lower 
in energy than  the other.
However, we should point out that they correspond to two 
different $\chi$ values and for the system to undergo a 
transition from the chiral restored to the chiral broken 
phase it has to go through a high potential energy barrier. 
In a 3D plot of the energy per baryon number versus $(\rho,\chi)$ 
the stable solutions correspond to two distinct
``valleys", and the unstable solution mentioned at the end of the 
previous section corresponds to the top of
 the barrier between the two valleys.

\section{Quark stars}

In order to 
investigate the structure of stars we solved the Tolman-Oppenheimer-Volkoff 
(TOV) equation  using the two EOS.
Since EOS-I is identical to the one using a quadratic potential, it leads to 
stars that have the same phenomenology as the hybrid stars obtained by Drago et al. \cite{drago}: 
$R\sim 10 - 12$ km, a hadron crust and a mass $M\sim 1- 2 M_\odot$. At 
 low densities, hadronization occurs and an hadronic equation 
of state should be used, replacing EOS-I. 

\begin{figure}[htb]
  \includegraphics[width=12cm]{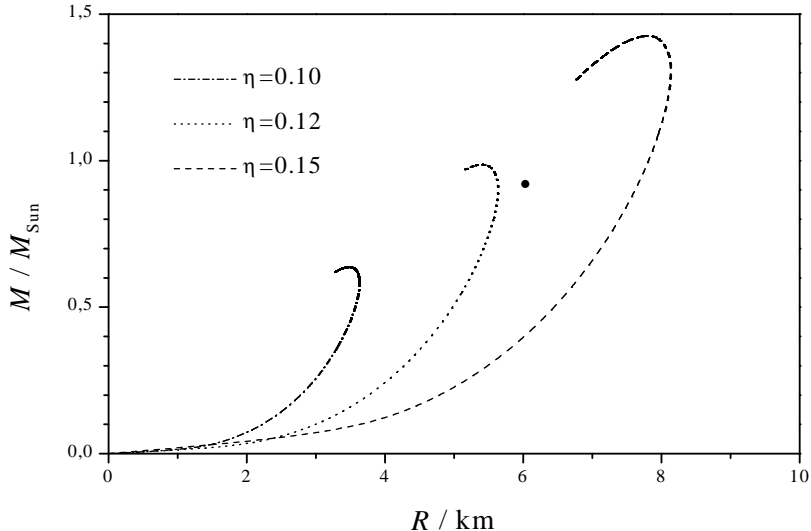}
\caption{Mass versus radius for the pure quark stars (solution II) 
in the CDM model. The dot indicates the maximum radius star for $\eta=0.115$. 
For the other model parameters see caption of Fig.~\ref{figstar3}.}
\label{figstar4}
\end{figure}

Since EOS-II saturates at a high density and, in addition, 
the system is not likely to undergo a transition to solution I,
one should not perform any connection to the hadronic sector: 
the EOS-II alone generates a new family of strange quark stars.  
In Fig.~\ref{figstar4} it is shown the mass-radius relation for 
different values of $\eta$.
These quark stars are smaller and 
denser in comparison with those resulting from EOS-I. For  $\eta\sim0.115$
(and $M=1.7$~GeV, yielding $B^{1/4}\sim 0.195$ GeV) one obtains 
a maximum radius $R\sim 6$~km and a
 corresponding mass $M\sim 0.9 M_\odot$, which are  
the fitted radius and mass for the nearby compact object 
RX J 1856.5-3754~\cite{pons,drake}. 
According to our calculation, such star has a central density of 
$10\rho_0$ ($\rho_0$ is the nuclear matter density) 
and a central energy density $\epsilon\sim 3\ttimes 10^{15}$ g/cm$^3$. 
At the edge, the density drops to $5\rho_0$
 and $\epsilon\sim 1.35\ttimes 10^{15}$ g/cm$^3$. 
The ratio $\epsilon/\rho$ remains approximately constant 
inside the star. The maximum period of the star, 
computed using the expression given in 
Ref.~\cite{haensel}, is $\sim0.4$~ms. 
From Fig.~\ref{figstar4} one concludes that the mass-radius relation for 
these strange small stars 
mainly depends on the height of the local minimum of the $\chi$ potential. 

\section{Conclusions}

 Using a mean-field  variational method we obtained two 
solutions for homogeneous strange matter in beta equilibrium using 
the CDM with a quartic potential, with its parameters fixed in the 
nucleon sector 
and to yield a reasonable bag constant. 

One solution is similar to the already 
known solution for quadratic potentials in the CDM, with massive quarks. 
In the other solution, quarks are almost   
massless. The EOS for both solutions are similar to those found recently 
in the framework of perturbative QCD. 

The pure quark stars emerging from 
the chiral symmetric 
solution are small and dense compact objects, showing a phenomenology 
compatible with 
the nearby X-ray source whose mass and radius was recently fitted~\cite{pons,drake}.  
 
\begin{theacknowledgments}

This work was supported by FCT (POCTI/FEDER program), Portugal and by CNPq/ICCTI through
the Brazilian-Portuguese scientific exchange program.  
We thank G. Marranghello and B. Garcia for some useful discussions.
E.O.A. and L.G.N. acknowledge the support of the PIBIC/CNPq program for young researchers. 
\end{theacknowledgments}

\end{document}